\documentstyle[prl,aps,epsfig]{revtex}

\newcommand{\beq}{\begin{equation}}
\newcommand{\eeq}{\end{equation}} 
\newcommand{\beqa}{\begin{eqnarray}}
\newcommand{\eeqa}{\end{eqnarray}}

\def\opone{\leavevmode\hbox{\small1\kern-3.8pt\normalsize1}}
 
\begin{document}

\draft

\twocolumn[\hsize\textwidth\columnwidth\hsize\csname
@twocolumnfalse\endcsname

\title{Quantum Trajectories for Brownian Motion}
\author
{Walter T.  Strunz$^{1}$, Lajos Di\'osi$^2$, Nicolas Gisin$^3$, 
and Ting Yu$^3$}
\address{
\protect\small\em $^1$Fachbereich Physik, Universit\"at GH Essen, 
45117 Essen, Germany\\
\protect\small\em $^2$Research Institute for Particle and Nuclear 
Physics, 1525 Budapest 114, POB 49, Hungary \\
\protect\small\em $^3$Group of Applied Physics, University of Geneva, 
1211 Geneva 4, Switzerland}
\date{\today}

\maketitle

\begin{abstract}
We present the stochastic Schr\"odinger equation for the dynamics of
a quantum particle coupled to a high temperature environment
and apply it to the dynamics of a driven, damped, nonlinear
quantum oscillator. Apart from an initial
slip on the environmental memory time scale, in the mean, our result 
recovers the solution of the known non-Lindblad quantum Brownian 
motion master equation. 
A remarkable feature of our powerful stochastic approach is its 
localization property:
individual quantum trajectories remain localized wave packets
for all times, even for 
the classically chaotic system considered here, the localization
being stronger as $\hbar \rightarrow 0$.
\end{abstract}

\pacs{03.65.Bz, 42.50.Lc, 05.40.+j}

\vskip2pc]
\narrowtext


The understanding of the dynamics of open or dissipative quantum
systems is of fundamental importance both from a practical and 
conceptual point of view.  The archetype of such a system is
the standard quantum Brownian motion model \cite{QBM} which describes 
a particle with Hamiltonian $H(q,p)$, coupled to an environment 
of harmonic oscillators $(q_\lambda,p_\lambda)$ 
via its position $q$, 
such that the total Hamiltonian of system and environment reads
\begin{eqnarray}\label{totalH}
H_{tot}(q,p,q_\lambda,p_\lambda)  & = &
H(q,p) + \\ \nonumber
& & \sum_\lambda\left\{\frac{p^2_\lambda}{2m_\lambda}
+\frac{1}{2}m_\lambda\omega_\lambda^2(q_\lambda - \frac{g_\lambda}
{m_\lambda\omega_\lambda^2} q)^2\right\}.
\end{eqnarray}

Up to now, in order to determine the time dependent 
dynamics of the open `system',
the standard procedure was the derivation of a master equation for the 
reduced density operator, which, for the high temperature case considered
below, is widely accepted to read
\begin{equation}\label{oldqbm}
\hbar \dot \rho_t = -i[H,\rho_t] - i\frac{\gamma}{2} [q,\{p,\rho_t\}] - 
\frac{m \gamma kT}{\hbar}[q,[q,\rho_t]],
\end{equation}
where $\gamma$ is the damping rate.
This master equation is a Markov master equation {\it not}, however, of
Lindblad form \cite{Lindblad}  and indeed it turns out that it may violate 
the positivity of $\rho_t$ on very short time scales, which has 
led to an ongoing debate about its range of applicability \cite{violation}. 
We will briefly address this issue later on in this Letter.

Our new approach to quantum Brownian motion is very different
and circumvents the derivation of a master equation for $\rho_t$
altogether. Instead, we use a stochastic Schr\"odinger
equation, derived straight from the microscopic model (\ref{totalH}),
for pure states $\psi_t(z)$ ({\it quantum trajectories}).
Our construction recovers the reduced density operator 
as the ensemble mean $M[\ldots]$ over many of these quantum 
trajectories, in principle without any approximation:
\begin{equation}\label{ensmean}
\rho_t = M\left[|\psi_t(z)\rangle\langle\psi_t(z)|\right].
\end{equation}
The mean $M[\ldots]$ is taken over the process $z_t$ which drives
the stochastic Schr\"odinger equation.
We are thus able to determine $\rho_t$ in a Monte-Carlo sense
without an explicit master equation for its time evolution.

Quantum trajectory methods have been used extensively in recent years,
mainly in the quantum optics community, due to their numerical
efficiency, their intimate connection to (continuous) measurement,
and their illustrative power helping to gain physical insight.
The master equations encountered in quantum optics are of standard 
Lindblad type, for which Markov
quantum trajectory methods are known for some time now: there are jump 
processes \cite{jumps} and diffusive processes \cite{QSD} recovering the
reduced density operator. Despite being maybe the best known of
all master equations, the Quantum Brownian motion master
equation (\ref{oldqbm}), being {\it not} of Lindblad form, has
so far been excluded from a treatment with these powerful methods.

Only recently the 
authors managed to extend the quantum trajectory concept
to {\it non-Markovian} situations \cite{NMQSD}, more precisely,
we were able to determine a stochastic Schr\"odinger equation for the
dynamics of a quantum system coupled to a bath
of harmonic oscillators as in (\ref{totalH}), without using
the concept of a master equation for $\rho_t$.
An alternative approach
to non-Markovian quantum trajectories, more emphasizing the
continuous measurement point of view, has now also been 
established \cite{ONMQSD}.

In its linear version \cite{LNMQSD},
our {\it non-Markovian quantum state diffusion} (QSD)
stochastic Schr\"odinger equation for the quantum Brownian motion 
model (\ref{totalH}) takes 
the form
\begin{equation}\label{LNMQSD}
\hbar \dot\psi_t(z) = -i H' \psi_t(z) + q z_t \psi_t(z)
-q\int_0^t\!ds\;\alpha(t,s) \frac{\delta\psi_t(z)}{\delta z_s},
\end{equation}
where we 
assumed a factorized total initial density operator 
$\rho_{tot} = |\psi_0\rangle\langle\psi_0|\otimes\rho_T$ with a pure system
state $|\psi_0\rangle$ and an environmental thermal density operator $\rho_T$.
The influence of the environment on the system is encoded in the bath
correlation function $\alpha(t,s) = \langle F(t) F(s)\rangle_{\rho_T}$
where $F(t) = \sum_\lambda g_\lambda q_\lambda(t)$ is the quantum force
in (\ref{totalH}) and $z_t$ is thus a complex 
Gaussian stochastic c-number force with correlation
$M[z^*_t z_s] = \alpha(t,s)$. In the usual high 
temperature limit $kT\gg\hbar\Lambda\gg\hbar\omega,\hbar\gamma$, where
$\Lambda$ is an environmental cutoff frequency and $\omega$, $\gamma$ are
the typical system frequency and damping rate, respectively, one finds
\cite{QBM}
\begin{equation}\label{hitalpha}
\alpha(t,s) = 2m \gamma kT\Delta(t-s) + i\hbar m\gamma\dot\Delta(t-s),
\end{equation}
where 
$\Delta(t)$ is a delta-like function decaying on the environmental
`memory' time scale $\Lambda^{-1}$ (here we use
$\Delta(t) = \frac{\Lambda}{2}e^{-\Lambda|t|}$).
In (\ref{LNMQSD}), the Hamiltonian
$H' = H(q,p) + \frac{1}{2}m\gamma\Lambda q^2$ 
contains an additional potential term that turns out to be
counterbalanced by a similar term arising from the memory integral.

Eq. (\ref{LNMQSD}) is exact, i.e. it provides a quantum trajectory
method for Brownian motion for any temperature and any distribution
of environmental oscillators in the model (\ref{totalH}), i.e.
for any $\alpha(t,s)$.
In order to compute numbers, however,
we have to express the functional derivative under the memory 
integral in (\ref{LNMQSD}) in terms of elementary operators. 
In the high temperature limit considered here, we simply need to
expand in terms of the time delay $(t-s)$
\begin{equation}\label{expand}
\frac{\delta\psi_t(z)}{\delta z_s} = \frac{1}{\hbar}
 \left(q - \frac{p}{m} (t-s) + \ldots \right)\psi_t(z),
\end{equation}
where the dots denote terms of the order $(t-s)^2$ and higher,
leading to corrections of the order $\omega/\Lambda$, $\gamma/\Lambda$ and
can therefore be neglected (see \cite{PertNMQSD} for a
general theory of such `post-Markov' open systems). 
With (\ref{expand}), the memory integral in
(\ref{LNMQSD}) takes the form
\begin{equation}\label{memint}
\int_0^t\! ds\, \alpha(t,s) 
\frac{\delta\psi_t(z)}{\delta z_s} = 
\left(g_0(t) q - g_1(t) p \right)\psi_t(z),
\end{equation}
where we introduce
time dependent coefficients
$g_0(t) = \frac{1}{\hbar} \int_0^t\!ds \alpha(t,s)$ and
$g_1(t) = \frac{1}{m\hbar}\int_0^t\!ds (t-s) \alpha(t,s).$
The imaginary part of $g_0(t)$ will be compensated by the additional
potential term in $H'$. The imaginary part of $g_1(t)$ gives rise to
damping. The real part of $g_0(t)$ describes diffusion and as the real
part of $g_1(t)$ also gives rise to diffusion, yet 
smaller by a factor $\omega/\Lambda$,
the latter can be neglected compared to the former in the regime
we are interested in.

In order to get an efficient Monte Carlo method
(importance sampling \cite{MonteCarlo}), we go over to the nonlinear
version of (\ref{LNMQSD}), which keeps the trajectories
$\psi_t(z)$ normalized at all times while retaining the correct
ensemble mean (\ref{ensmean}), see \cite{NMQSD}. Using (\ref{memint}),
the relevant
stochastic Schr\"odinger equation for Brownian motion reads
\begin{eqnarray}\label{QSDQBM}
\hbar\dot\psi_t(z) & = & -iH\psi_t(z) -i\left(\frac{1}{2}m\gamma\Lambda+
\mbox{Im}\{g_0(t)\}\right)q^2 \psi_t(z)\\
\nonumber
 & & + (q-\langle q\rangle)z_t\psi_t(z) \\
\nonumber
 & & - \mbox{Re}\{g_0(t)\}
\left((q-\langle q\rangle)^2 - \langle(q-\langle q\rangle)^2\rangle
\right) \psi_t(z) \\ 
\nonumber
 & & +i\mbox{Im}\{g_1(t)\}\left( qp-\langle qp\rangle + 
m\dot{\langle q\rangle} q 
- \langle q\rangle p \right) \psi_t(z).
\end{eqnarray}
Normalized quantum trajectories $\psi_t(z)$
whose ensemble mean gives the desired reduced density operator
according to (\ref{ensmean}) can now be propagated using (\ref{QSDQBM}),
where $\dot{\langle q\rangle} = \frac{d}{dt}\langle q\rangle$, a quantity
which has to be determined numerically along with $\psi_t(z)$
(very often the replacement 
$m\dot{\langle q\rangle} \approx \langle p\rangle$ turns
 out be a good approximation).

In (\ref{QSDQBM}), the time dependent coefficients
quickly approach their asymptotic values
$g_0(t) \rightarrow \frac{m \gamma kT}{\hbar} - 
\frac{i}{2}m\gamma\Lambda$ 
and Im$\{g_1(t)\} \rightarrow -\frac{\gamma}{2}$ for
times larger than the environmental memory time.
After this {\it initial slip}
$t\gg\Lambda^{-1}$, (\ref{QSDQBM}) becomes
\begin{eqnarray}\label{QSDQBMasy}
\hbar\dot\psi_t(z) & = & -iH\psi_t(z) 
  + (q-\langle q\rangle)z_t\psi_t(z) \\
\nonumber
 & & - \frac{m\gamma kT}{\hbar}
\left((q-\langle q\rangle)^2 - \langle(q-\langle q\rangle)^2\rangle
\right) \psi_t(z) \\ 
\nonumber
& & -\frac{i}{2}\gamma\left( qp-\langle qp\rangle + 
m\dot{\langle q\rangle} q -
                     \langle q\rangle p \right) \psi_t(z).
\end{eqnarray}

We now highlight the power of our stochastic Schr\"odinger equation
for Brownian motion (\ref{QSDQBM}) by investigating the dynamics of a
driven, damped, nonlinear, noisy system, the Duffing
oscillator, where $H = \frac{1}{2}p^2 + \frac{1}{4}q^4 - \frac{1}{2}q^2
+ g q \cos(t)$, here coupled to a heat bath at temperature $T$.
This system has been studied before using the master equation
(\ref{oldqbm}) (see \cite{Zurek} and references therein),
including a straight numerical solution
which requires the propagation of a huge matrix.
In our new approach, one propagates pure states $\psi_t(z)$
according to (\ref{QSDQBM}), a great reduction in resources,
with the need, however, to solve (\ref{QSDQBM})  many times
in order to evaluate the mean values. For Lindblad master equations,
the power of quantum trajectory methods for investigating
classically chaotic dissipative systems was shown in \cite{SpillerRalph}
(see also \cite{QSDlocalization}).

We use parameters $g=0.3$ with a damping rate $\gamma = 0.25$,
thus the classical problem is chaotic \cite{nonlinbook}.
The environment is furthermore characterized by
$kT = 0.3$, and a cutoff frequency $\Lambda=5$.
With $\hbar$ of the order $10^{-2}$ and smaller 
(see various choices of $\hbar$
below), the parameters are in the required regime.
As initial condition we choose a standard
coherent state located at 
$\langle q \rangle = 0.1, \langle p \rangle = 0.1$.

In Fig.1 we show the ensemble mean $M[W_z(q,p,t=4)]$ 
over $1000$, $5000$, and $10000$ Wigner functions 
of pure state trajectories $\psi_t(z)$ obtained solving 
(\ref{QSDQBM}) numerically up to a time $t=4$.
According to our construction, this quantity 
converges to the Wigner function of the reduced density operator
for many realizations.
Here we have chosen $\hbar = 0.01$, a phase space
area corresponding
approximately to the extension of the wave packets shown in Fig.2.

\centerline{\epsfig{figure=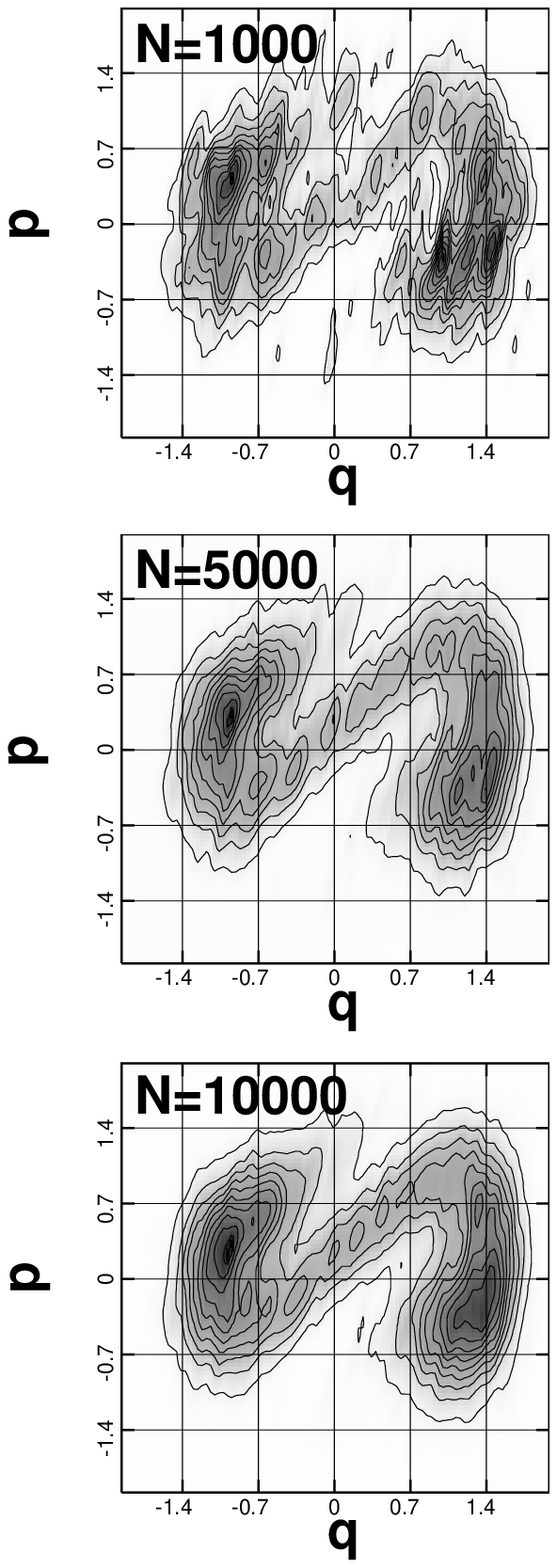,height=17cm,angle=0}}
{FIG. 1. {\small{
Contour plots of the Wigner function $W(q,p,t=4)$
 of the reduced density operator of the 
thermal Duffing oscillator with $\hbar=0.01$ (for the
phase space area corresponding to this $\hbar$ see Fig.2). 
The contour plots show the ensemble mean over $1000$, $5000$, 
and $10000$ Wigner functions $W_z(q,p,t=4)$
of individual quantum trajectories obtained solving the
quantum Brownian motion stochastic Schr\"odinger equation 
(\ref{QSDQBM}).  }}}\\

In Fig.2 we show contour plots of Wigner functions $W_z(q,p,t=4)$ 
of four realizations
of (\ref{QSDQBM}), many of which add up to the Wigner function of
the desired reduced density matrix shown in Fig.1. 
One can see clearly that these individual
Wigner functions are well localized in phase space compared to the
phase space spread of the ensemble, even for this classically chaotic
system. 

\centerline{\epsfig{figure=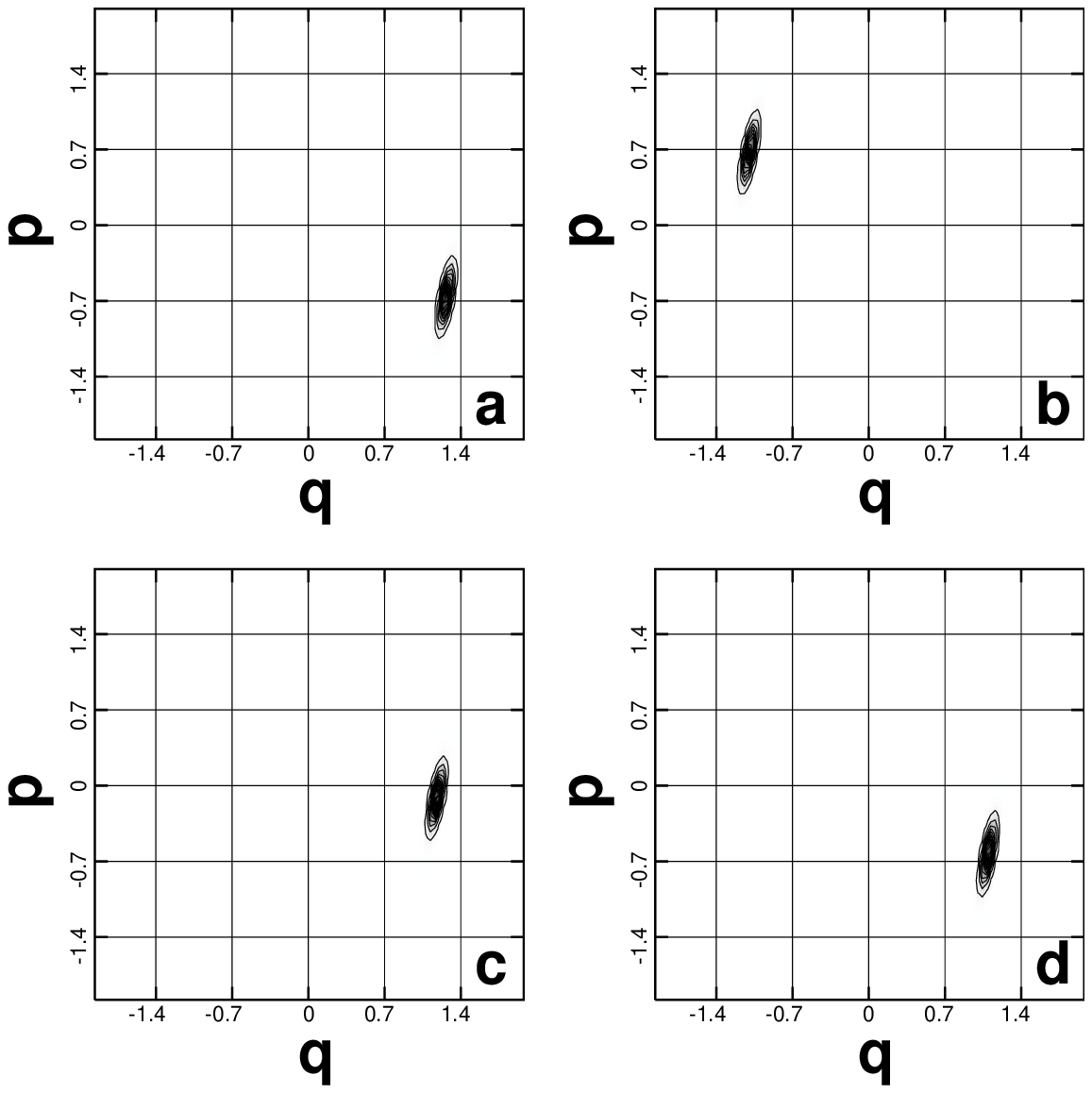,height=6cm,angle=0}}
{FIG. 2. {\small{
Contour plots of Wigner functions $W_z(q,p,t=4)$ of
four individual quantum trajectories obtained solving
the quantum Brownian motion stochastic Schr\"odinger equation
(\ref{QSDQBM}) for the thermal Duffing oscillator. 
Individual trajectories remain well localized in phase space with
respect to the overall spread of the ensemble mean, even
for this classically chaotic system. The chosen value of $\hbar=0.01$
is slightly smaller than the phase space area covered by these states.
}}}\\

This remarkable feature of the quantum Brownian motion
stochastic Schr\"odinger equation (\ref{QSDQBM}) is highlighted
again in Fig.3, where we show the mean position 
spread, $M[\Delta q] = M[\sqrt{\langle(q-\langle q\rangle)^2\rangle}]$
and the mean uncertainty product
$M[\Delta q \Delta p/\hbar]$ in units of $\hbar$ of individual 
trajectories as a function of time for three different choices of
$\hbar$.

\centerline{\epsfig{figure=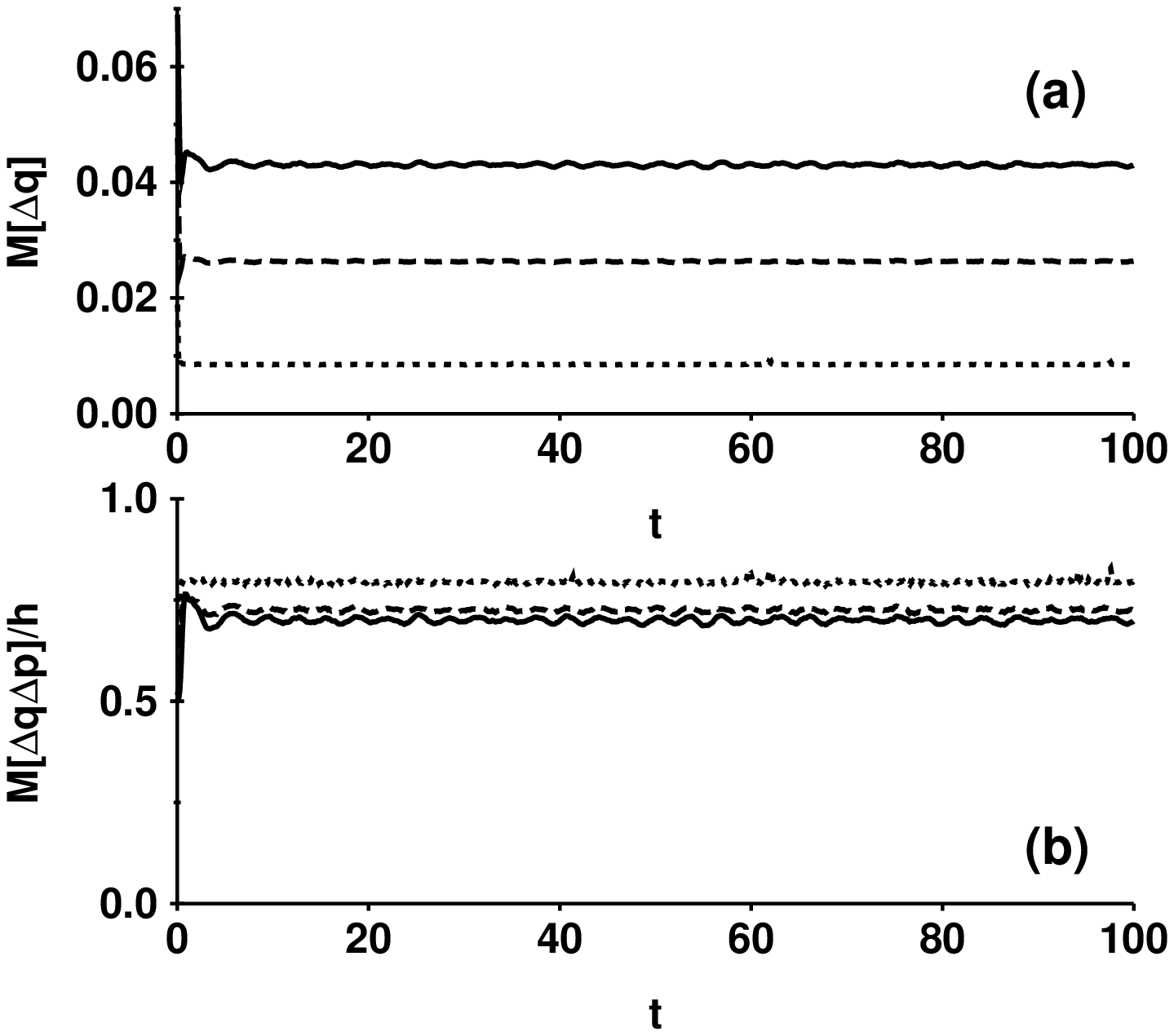,height=5.5cm,angle=0}}
{FIG. 3. {\small{
Localization property of the QBM stochastic Schr\"odinger equation.
Individual runs are well localized in phase space, the localization
being stronger the smaller $\hbar$: (a) the
average position spread
$M[\Delta q] = M[\sqrt{\langle(q-\langle q\rangle)^2\rangle}]$
of solutions of the QBM stochastic Schr\"odinger equation
for the choices $\hbar=0.01$ (solid line),
$\hbar=0.005$ (dashed line), and $\hbar = 0.001$ (dotted line).
Fig. (c) shows the mean uncertainty product
$M[\Delta q\Delta p]/\hbar$, which remains of the order one
almost independently of $\hbar$.
Thus the quantum trajectories remain almost minimum uncertainty 
wave packets for all times.
}}}\\

The quantities shown in Fig.3 can only be given sense in the 
framework of quantum trajectories, they
have no meaning from a density operator point of view as they are 
the ensemble mean over an expression non-quadratic in $\psi_t(z)$.
It is apparent from Fig.3 that individual trajectories are well 
localized in phase space for all times, the localization
being stronger the smaller $\hbar$. 
As can be seen, our quantum trajectories remain
almost `classical' states,
yet recover the fully quantum master equation (\ref{oldqbm}).
Thus, the representation (\ref{ensmean})
expresses the reduced density operator of quantum
Brownian motion explicitly as a mixture of almost `classical' states.

The observed localization property of QSD is well known
in the Markov case and has been studied for instance in 
\cite{QSDlocalization}. 
Here we see that similar properties hold for the generalized
non-Markovian QSD equation (\ref{QSDQBM})
which has now been applied to
quantum dynamics beyond the class of Lindblad master equations.
As in the Markov case, the localization property can be exploited 
to further reduce the numerical effort.

Finally, let us briefly address the connection between our approach 
and the widely used QBM master 
equation (\ref{oldqbm}). Since a quantum trajectory approach strictly
preserves positivity of the reduced density operator, our
QBM stochastic Schr\"odinger equation (\ref{QSDQBM}) cannot be identical 
to (\ref{oldqbm}) in the mean, as the latter is known to violate positivity
on short time scales.
Taking the ensemble mean $M\left[\ldots\right]$ in (\ref{ensmean})
with (\ref{LNMQSD})
analytically, we were able to show in \cite{PertNMQSD} that in the regime
considered in this Letter, the evolution of the ensemble mean 
(\ref{ensmean}) is well described by
the master equation
\begin{eqnarray}\label{newqbm}
\hbar \dot \rho & = & -i[H,\rho] - i\left(\frac{1}{2}m\gamma\Lambda 
+ \mbox{Im}
\{g_0(t)\}\right) [q^2,\rho] \\ \nonumber & &
+i\mbox{Im}\{g_1(t)\} [q,\{p,\rho\}] - \mbox{Re}\{g_0(t)\}[q,[q,\rho]],
\end{eqnarray}
which reduces to (\ref{oldqbm}) for times larger than the environmental
memory time, $t\gg\Lambda^{-1}$ 
due to the asymptotics of the coefficients $g_0(t), g_1(t)$. Thus, 
apart from an initial slip on the environmental memory time scale
$\Lambda^{-1}$, 
our approach recovers (\ref{oldqbm}) in the mean.
It is known in the case of the exact master equation
for a damped harmonic oscillator \cite{HuPaz} that such time dependent
coefficients may ensure the positivity of the reduced density
operator  for non-Lindblad master equations, a result that
is here supported for general system Hamiltonian $H(q,p)$.

To conclude, we have presented the stochastic Schr\"odinger equation
for Brownian motion. It is compatible with the standard QBM
master equation yet allows to compute states rather than a matrix,
a huge reduction in resources,
which becomes even more relevant for QBM in more than one space 
dimension. Individual trajectories are well localized in phase space,
the localization being stronger the smaller $\hbar$. 
Thus, in (\ref{ensmean}), 
the reduced density operator for Brownian motion
is explicitly represented as an ensemble of almost `classical' states.

We thank F Haake and IC Percival for helpful comments.
WTS would like to thank
the Deutsche Forschungsgemeinschaft for support through the SFB 237
``Unordnung und gro{\ss}e Fluktuationen''.
NG and TY thank the Swiss National Science Foundation.

\end{document}